\documentclass[twocolumn,showpacs,amssymb,aps,superscriptaddress,pre]{revtex4}
\usepackage{graphicx}
\usepackage{dcolumn}
\usepackage{bm}

\begin{document}

\title{Effect of on-site Coulomb repulsion on phase transitions in exactly solved spin-electron model}
\author{L. G\'ALISOV\'A}
\affiliation{Department of Applied Mathematics, Faculty of Mechanical Engineering, \\
Technical University, Letn\'a 9, 042 00 Ko\v{s}ice, Slovak Republic}
\author{J. STRE\v{C}KA}
\affiliation{Department of Theoretical Physics and Astrophysics, Faculty of Science, \\
P. J. \v{S}af\'{a}rik University, Park Angelinum 9, 040 01 Ko\v{s}ice, Slovak Republic}
\author{A. TANAKA}
\affiliation{Department of General Education, Ariake National College of Technology,
Omuta, Fukuoka 836-8585, Japan}
\author{T. VERKHOLYAK}
\affiliation{Institute for Condensed Matter Physics, National Academy of Sciences of Ukraine,
1 Svientsitskii Street, L'viv-11, 790 11, Ukraine}

\begin{abstract}
A hybrid lattice-statistical model on doubly decorated planar lattices, which have localized Ising spins at their nodal lattice sites and two itinerant electrons at each pair of decorating sites, is exactly solved by the use of a generalized decoration-iteration transformation. Our main attention is focused on an influence of the on-site Coulomb repulsion on ground-state properties and critical behavior of the investigated system.
\end{abstract}
\pacs{05.50.+q; 05.70.Jk; 64.60.Cn; 64.60.De}

\maketitle

\section{Introduction}
Exactly soluble lattice-statistical models attract scientific interest as they offer a valuable insight into diverse aspects of cooperative phenomena~\cite{wu09}. The mapping technique based on generalized algebraic transformations~\cite{fis59} belongs to the simplest mathematical methods, which allow to obtain the exact solution for various statistical models. Recently, this approach has been applied to an interesting diamond-chain model of interacting spin-electron system~\cite{per09}. This motivated us to start studying a similar spin-electron system on doubly decorated two-dimensional (2D) lattices in order to provide a deeper insight into how the mobile electrons influence their magnetic properties~\cite{str09}. The investigated model system might provide guidance on a magnetic behavior of magnetic metals such as SrCo$_6$O$_{11}$~\cite{ish05}, which contain both localized spins and itinerant electrons.

\section{Model and its exact solution}
Consider a hybrid lattice-statistical model of interacting spin-electron system on doubly decorated 2D lattices (see Fig.~1 in Ref.~\cite{str09}), which have one localized Ising spin at each nodal lattice site and two delocalized mobile electrons at each couple of decorating sites. The total Hamiltonian of the model can be written as a sum over bond Hamintonians $\hat{\cal H} = \sum_{k}\hat{\cal H}_k$, where each bond Hamiltonian $\hat{\cal H}_k$ involves all interaction terms of $k$th couple of delocalized electrons
\begin{eqnarray}
\hat{\cal H}_k \!\!&=&\!\! -t\sum_{\alpha = \uparrow,\downarrow}\left( c^{\dagger}_{k1,\alpha}c_{k2,\alpha}+{\rm h.c.} \right)
+ U \sum_{i=1}^2 n_{ki,\uparrow}n_{ki,\downarrow}
\nonumber \\
\!\!& &\!\!
-\frac{J}{2} \sum_{i=1}^2\hat{\sigma}^z_{ki}\left(c^{\dagger}_{ki,\uparrow}c_{ki,\uparrow} - c^{\dagger}_{ki,\downarrow}c_{ki,\downarrow}\right).
\label{2.01}
\end{eqnarray}
Here, $c^{\dagger}_{ki,\alpha}$ and $c_{ki,\alpha}$ ($\alpha = \uparrow,\downarrow$) represent usual creation and annihilation fermionic operators, respectively, $n_{i,\alpha} = c^{\dagger}_{ki,\alpha}c_{ki,\alpha}$ and $\hat{\sigma}^z_{ki}$ is the standard spin-1/2 operator. The hopping parameter $t$ takes into account kinetic energy of mobile electrons, $J$ describes the Ising interaction between the itinerant electrons and their nearest Ising neighbors and $U\geq0$ denotes the on-site Coulomb repulsion. Following the rigorous approach developed in Ref.~\cite{str09}, one obtains a simple universal equality
\begin{eqnarray}
{\cal Z}(T, J, t, U) = A^{Nq/2}{\cal Z}_{\rm Ising}(T, R),
\label{2.02}
\end{eqnarray}
which establishes an exact mapping relationship between the partition function $\cal{Z}$ of the interacting spin-electron system
on the doubly decorated 2D lattice and the partition function $\cal{Z}_{\rm Ising}$ of the spin-1/2 Ising model on
the corresponding undecorated lattice of $N$ vertices with the coordination number $q$ and the effective temperature-dependent nearest-neighbor interaction $R$. Note that the relation (\ref{2.02}) allows direct calculation of all relevant physical quantities to be useful for understanding of magnetic behavior of the considered system.

\section{Results and Discussion}
\begin{figure}[htb]
\vspace{-1.0cm}
\begin{center}
\includegraphics[width=7.75cm]{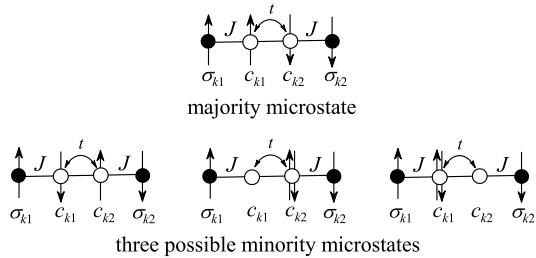}
\end{center}
\vspace{-0.9cm}
\caption{Four entangled microstates, which emerge in $|{\rm AF}\rangle$.}
\label{fig1}
\end{figure}

Now, let us proceed to a discussion of the most interesting numerical results obtained for the interacting spin-electron model on doubly decorated 2D lattices with the on-site Coulomb repulsion. Before doing this, however, it is noteworthy that the special case of this hybrid system without the on-site repulsion term has already been examined in detail by present authors in the earlier paper~\cite{str09}. The exact results for this particular model revealed that the ground state of the system consists of an interesting four-sublattice quantum antiferromagnetic phase $|{\rm AF}\rangle$. $|{\rm AF}\rangle$ is characterized by a perfect N\'eel order of the Ising spins situated at the nodal lattice sites and the quantum entanglement of four microstates $c^{\dagger}_{k1,\uparrow}c^{\dagger}_{k2\downarrow}|0\rangle$, $c^{\dagger}_{k1,\downarrow}c^{\dagger}_{k2,\uparrow}|0\rangle$, $c^{\dagger}_{k1,\uparrow}c^{\dagger}_{k1,\downarrow}|0\rangle$, and $c^{\dagger}_{k2,\uparrow}c^{\dagger}_{k2,\downarrow}|0\rangle$ of itinerant electrons delocalized over decorating lattice sites (see Fig.~\ref{fig1}), whose probability amplitudes depend on a mutual ratio between the hopping term $t$ and the Ising interaction $J$.
With this background, the purpose of the present work is to shed light on how the on-site Coulomb repulsion $U>0$ affects the spontaneous long-range order to emerge in $|{\rm AF}\rangle$. For simplicity, we will further assume the ferromagnetic Ising interaction $J>0$, because the sign change $J \to -J$ brings just a trivial change in the mutual orientation of the nodal Ising spins with respect to their nearest-neighbor electrons.

\begin{figure}[ht]
\vspace{-1.0cm}
\begin{center}
\includegraphics[width=7cm]{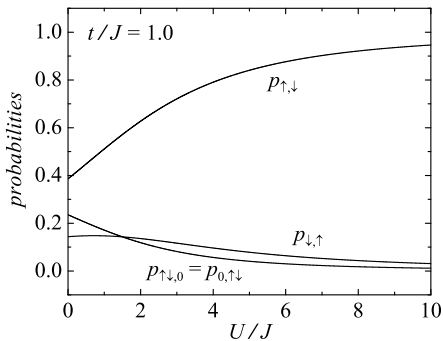}
\end{center}
\vspace{-1.0cm}
\caption{The probability distribution of the entangled microstates as a function of the on-site repulsion for $t/J=1.0$.}
\label{fig2}
\end{figure}

The deeper insight into the effect of the on-site Coulomb repulsion on a probability distribution of four entangled microstates in $|{\rm AF}\rangle$ may be gained from Fig.~\ref{fig2}, which shows the abundance probabilities of relevant microstates as functions of $U/J$ for the fixed $t/J = 1.0$.
As one can clearly see from this figure, the abundance probability $p_{\uparrow,\downarrow}$ of the majority microstate $c^{\dagger}_{k1,\uparrow}c^{\dagger}_{k2\downarrow}|0\rangle$ monotonically increases upon strengthening the on-site repulsion at the expense of the abundance probabilities $p_{\downarrow,\uparrow}$, $p_{\uparrow\downarrow,0}$, $p_{0,\uparrow\downarrow}$ of minority microstates, until they asymptotically reach the values $p_{\uparrow,\downarrow} \to 1$ and $p_{\downarrow,\uparrow} = p_{\uparrow\downarrow,0} = p_{0,\uparrow\downarrow} \to 0$, respectively, in the limit $U/J \to \infty$. Consequently, one may conclude that in the limit of infinitely strong on-site repulsion each decorating lattice site is occupied by just one mobile electron, which prefers the ferromagnetic alignment with respect to its nearest Ising neighbor. Providing that within the second-order perturbation theory the kinetic term $t$ at half-filling is equivalent to the AF Heisenberg interaction between spins, one may conclude that for sufficiently strong repulsion, the ground state of the interacting spin-electron system is identical with the one of the spin-1/2 Ising-Heisenberg antiferromagnet defined on the same doubly decorated lattice~\cite{jas02}.
\begin{figure}[htb]
\vspace{-1.0cm}
\begin{center}
\includegraphics[width=7cm]{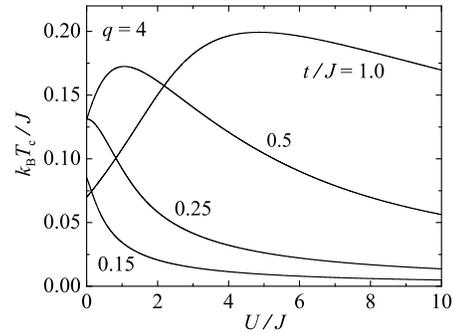}
\end{center}
\vspace{-1.0cm}
\caption{The critical temperature of the interacting spin-electron system on the doubly decorated square lattice as a function of $U/J$ for several values of the kinetic term $t/J$.}
\label{fig3}
\end{figure}
In Fig.~\ref{fig3}, we depict the critical temperature of the spin-electron model on doubly decorated square lattice as a function of the the on-site Coulomb repulsion for several values of the kinetic term. As one can see, the critical temperature monotonically decreases upon strengthening the Coulomb repulsion for sufficiently small hopping terms $t/J \lesssim 0.25$, while for stronger kinetic terms $t/J \gtrsim 0.25$
it exhibits a more striking dependence with a pronounced maximum due to a mutual competition between the sufficiently strong hopping term $t/J$ and
the Coulomb term $U/J$.

In conclusion, the effect of on-site Coulomb repulsion on the ground-state properties and critical behavior of the interacting spin-electron system on doubly decorated 2D lattices have been examined using the generalized decoration-iteration transformation. The obtained exact results clearly show that the dependence of critical temperature on a strength of the on-site Coulomb repulsion basically depends on whether the kinetic term is greater than or less than the boundary value $t/J \thickapprox 0.25$.

{\bf Acknowledgments}:
L.G. and J.S. acknowledge financial support under the grant VEGA 1/0431/10.

\end{document}